\documentclass[conference]{IEEEtran}
\IEEEoverridecommandlockouts
\usepackage{cite}
\usepackage{amsmath,amssymb,amsfonts}
\usepackage{algorithmic}
\usepackage{graphicx}
\usepackage{textcomp}
\usepackage{xcolor}
\def\BibTeX{{\rm B\kern-.05em{\sc i\kern-.025em b}\kern-.08em
    T\kern-.1667em\lower.7ex\hbox{E}\kern-.125emX}}
    
\usepackage{enumitem}
\usepackage{caption}
\usepackage{comment}
\usepackage{xcolor}
\usepackage{amsfonts}
\usepackage{ragged2e}
\usepackage{amsmath}
\usepackage{multirow}

\newcommand{\etcno}{\textit{etc}}
\newcommand{\egno}{\textit{e}.\textit{g}.} 
\newcommand{\ieno}{\textit{i}.\textit{e}.}

\begin{document}

\title{Frequency-Assisted Adaptive Sharpening Scheme Considering Bitrate and Quality Tradeoff
}

\author{\IEEEauthorblockN{1\textsuperscript{st} Yingxue Pang}
\IEEEauthorblockA{
\textit{Bytedance Inc.}\\
Shanghai, China \\
pangyingxue@bytedance.com}
\IEEEauthorblockN{4\textsuperscript{th} Gen Zhan}
\IEEEauthorblockA{
\textit{Bytedance Inc.}\\
Shenzhen, China \\
zhangen@bytedance.com}
\and
\IEEEauthorblockN{2\textsuperscript{nd} Shijie Zhao*}
\thanks{* Corresponding author}
\IEEEauthorblockA{
\textit{Bytedance Inc.}\\
Shenzhen, China \\
zhaoshijie.0526@bytedance.com}
\IEEEauthorblockN{5\textsuperscript{th} Junlin Li}
\IEEEauthorblockA{
\textit{Bytedance Inc.}\\
San Diego, CA, 92122 USA \\
lijunlin.li@bytedance.com}
\and
\IEEEauthorblockN{3\textsuperscript{rd} Haiqiang Wang}
\IEEEauthorblockA{
\textit{Bytedance Inc.}\\
Shenzhen, China \\
wanghaiqiang@bytedance.com}
\IEEEauthorblockN{6\textsuperscript{th} Li Zhang}
\IEEEauthorblockA{
\textit{Bytedance Inc.}\\
San Diego, CA, 92122 USA \\
lizhang.idm@bytedance.com}
}

\maketitle

\begin{abstract}
Sharpening is a widely adopted technique to improve video quality, which can effectively emphasize textures and alleviate blurring. However, increasing the sharpening level comes with a higher video bitrate, resulting in degraded Quality of Service (QoS). Furthermore, the video quality does not necessarily improve with increasing sharpening levels, leading to issues such as over-sharpening. Clearly, it is essential to figure out how to boost video quality with a proper sharpening level while also controlling bandwidth costs effectively. This paper thus proposes a novel \textbf{Freq}uency-assisted \textbf{S}harpening level \textbf{P}rediction model (\textbf{FreqSP}). We first label each video with the sharpening level correlating to the optimal bitrate and quality tradeoff as ground truth. Then taking uncompressed source videos as inputs, the proposed FreqSP leverages intricate CNN features and high-frequency components to estimate the optimal sharpening level. Extensive experiments demonstrate the effectiveness of our method.
\end{abstract}


\section{Introduction}
\label{sec:intro}
With the advancement of digital media and hardware, various types of video traffic such as digital television broadcasting, Video-on-Demand (VoD), Internet video streaming and P2P have become increasingly popular. 
Thus, it has become imperative for video service providers to ensure the delivery of videos with satisfactory quality.
However, acquiring high-quality video can be quite challenging due to the numerous distortions that occur in the encoding, storing and streaming processes on communication networks. As such, enhancement operations are necessary to improve low-quality video to a more acceptable standard. 
One of the most reliable and widely used enhancement techniques is sharpening, which can effectively improve video quality by emphasizing texture, overcoming blurring, and drawing the attention of viewers to certain areas.

\begin{figure}[!t]
\centering
\includegraphics[width=0.5\textwidth]{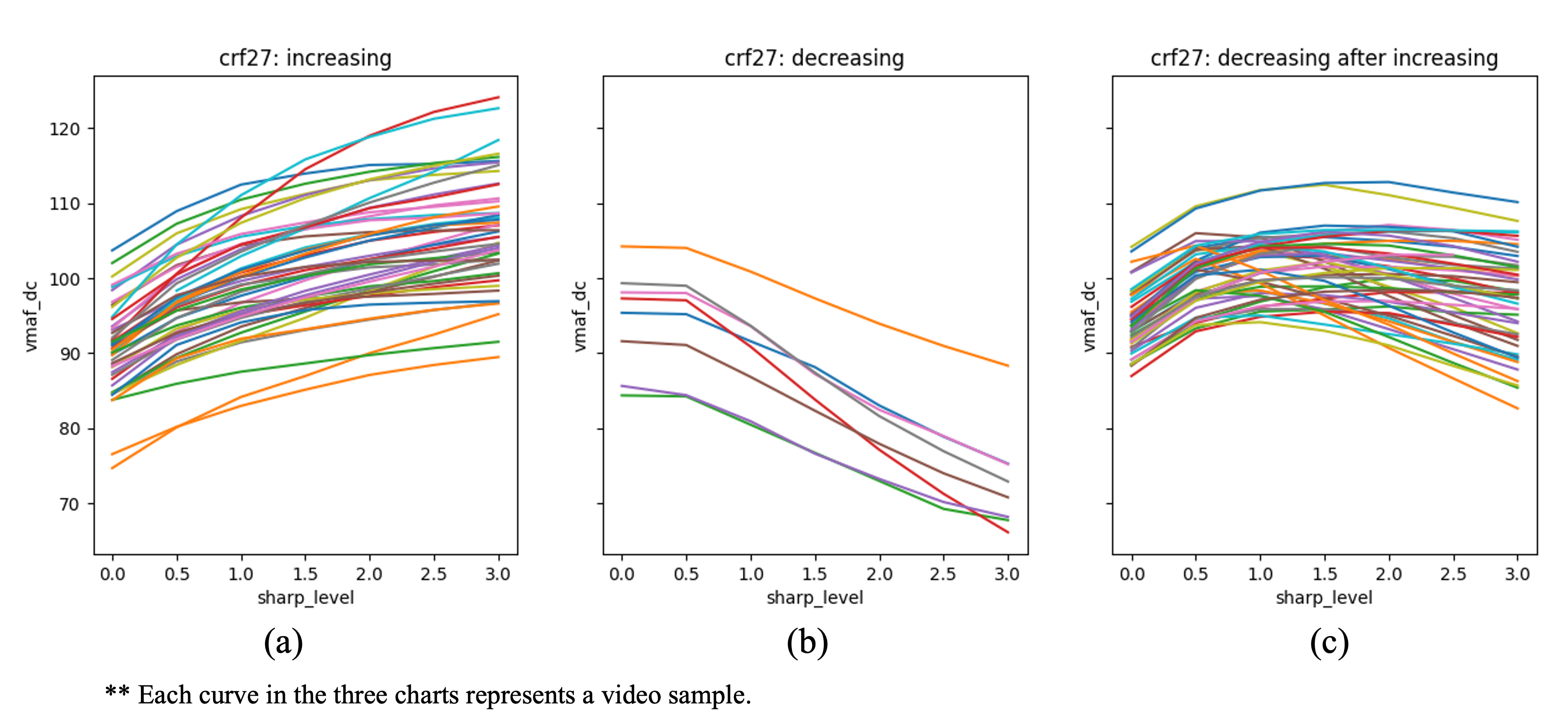}
\caption{Relationship between sharpening level and quality.}
\label{fig:motivation}
\end{figure}

Empirically, it has been observed that an increase in sharpening levels does not necessarily result in the improvement of video quality. To further verify this, we randomly sample 100 videos from LSVQ~\cite{ying2021patch} and used FFmpeg's built-in USM function~\cite{jain1989fundamentals} to sharpen them to varying degrees. Afterwards, these sharpened videos are compressed using the HEVC/H.265 codec~\cite{sullivan2012overview} with CRF 27. The relationship between the sharpening level and video quality is then visualized. As illustrated in Fig.\ref{fig:motivation} (a), an increase in the sharpening level improves video quality. However, the opposite can be seen in Fig.\ref{fig:motivation} (b) where the quality of the video decreases as the sharpening level increases. Fig.\ref{fig:motivation} (c) shows that video quality increases initially before declining as the sharpening level is increased. 
These observations indicate that applying a higher sharpening intensity can potentially lead to poorer video quality or a lower Quality of Experience (QoE). More importantly, as the sharpening level increases, so does the video bitrate, driving up bandwidth costs and degrading Quality of Service (QoS) that manifests in the form of buffering, jitter and first-frame delay, \etcno. In other words, there are instances when we use higher bandwidth costs while suffering unsatisfactory QoE and degraded QoS.

\begin{figure*}[!t]
\centering
\includegraphics[width=\textwidth]{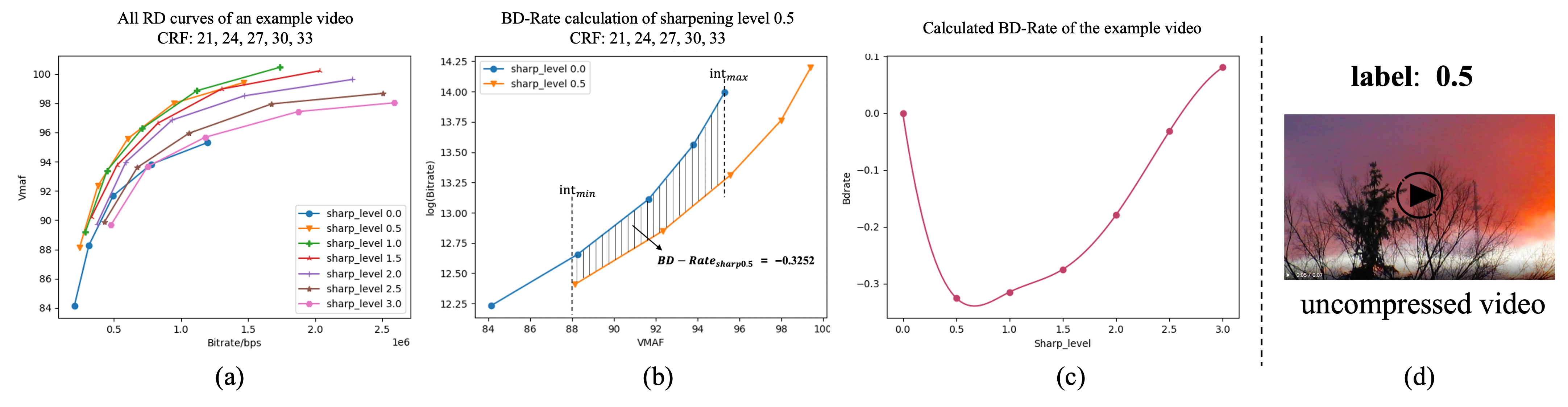}
\caption{Illustration of the pseudo-labeling process with BD-Rate. (a) shows all RD curves of an example video. (b) shows the BD-Rate calculation of sharpening level 0.5. (c) shows all the calculated BD-Rate values of different sharpening levels. (d) Sharpening level 0.5 has the minimal BD-Rate value, so we label this example video as 0.5.}
\label{fig:label}
\end{figure*}

To address the issues mentioned above, we believe it is imperative to determine the optimal sharpening level to improve video quality while efficiently saving bandwidth costs. In more detail, we propose a novel \textbf{Freq}uency-assisted \textbf{S}harpening level \textbf{P}rediction model (\textbf{FreqSP}) that uses pseudo-labeled uncompressed source videos as input to estimate the ideal sharpening level. Each training video is pseudo-labeled using the sharpening level associated with the optimal bitrate and quality tradeoff. We use the Bjøntegaard-Delta bitrate (BD-Rate)~\cite{bjontegaard2001calculation} to measure the aforementioned tradeoff derived from its Rate-Distortion (RD) characteristics. After labeling, our proposed FreqSP fuses the intricate CNN features and high-frequency components extracted from input uncompressed source videos to predict the BD-Rate-related sharpening level.

The main contributions of this paper are summarized as follows:
\begin{itemize}
	\item We propose a novel \textbf{Freq}uency-assisted \textbf{S}harpening level \textbf{P}rediction model (\textbf{FreqSP}) to fuse CNN features and high-frequency components to estimate the optimal sharpening level with uncompressed videos as input.
	
	\item To the best of our knowledge, FreqSP is the first sharpening level prediction model trained on BD-Rate-related pseudo-label considering the optimal bitrate and quality tradeoff. The predicted sharpening level can effectively improve video quality while saving unnecessary bandwidth.
	
	\item Extensive experiments on multiple benchmarks demonstrate the effectiveness of our method in terms of various quantitative metrics.
	
\end{itemize}

\section{Frequency-Assisted Adaptive Sharpening Scheme}
\label{sec:method}

In this section, we first introduce the labeling paradigm which aims to assign an optimal BD-Rate-related sharpening level to each uncompressed training video as ground truth (Section~\ref{sec:labelling}). Afterwards, the uncompressed videos with the relevant assigned labels are utilized as inputs for the proposed Frequency-assisted Sharpening level Prediction model (FreqSP, Section~\ref{sec:pred}) to carry out the training process.

\subsection{Pseudo optimal sharpening level labeling}
\label{sec:labelling}

Our model is designed to predict the optimal sharpening level to improve video quality and reduce bandwidth consumption. It is crucial to ensure that the assigned label accurately reflects the tradeoff between quality and bitrate. In other words, the video sharpened with the assigned sharpening level should provide the maximum perceptual quality gain with the fewest bits after compression. Consequently, the first and most important issue is how to acquire these ideal sharpening levels.

The performance analysis approaches employed in video coding draw our attention. RD curves are used to illustrate how well an encoder performs, with higher quality (\egno, PSNR, SSIM, VMAF, \etcno) for lower bitrates indicating better encoder performance. And BD-Rate is used to assess the compression efficiency of an encoder compared to a reference encoder, also known as an anchor, by calculating the average quality difference between two RD curves over an interval. Here, we define each encoder as a paradigm of sharpening at different levels and then encoding at different CRFs. We use the encoder with sharpening level 0.0 as an anchor to calculate the BD-Rate for the other sharpening levels. The pseudo-label for training is then assigned as the sharpening level with the biggest BD-Rate gain.

\begin{figure*}[!t]
\centering
\includegraphics[width=1\textwidth]{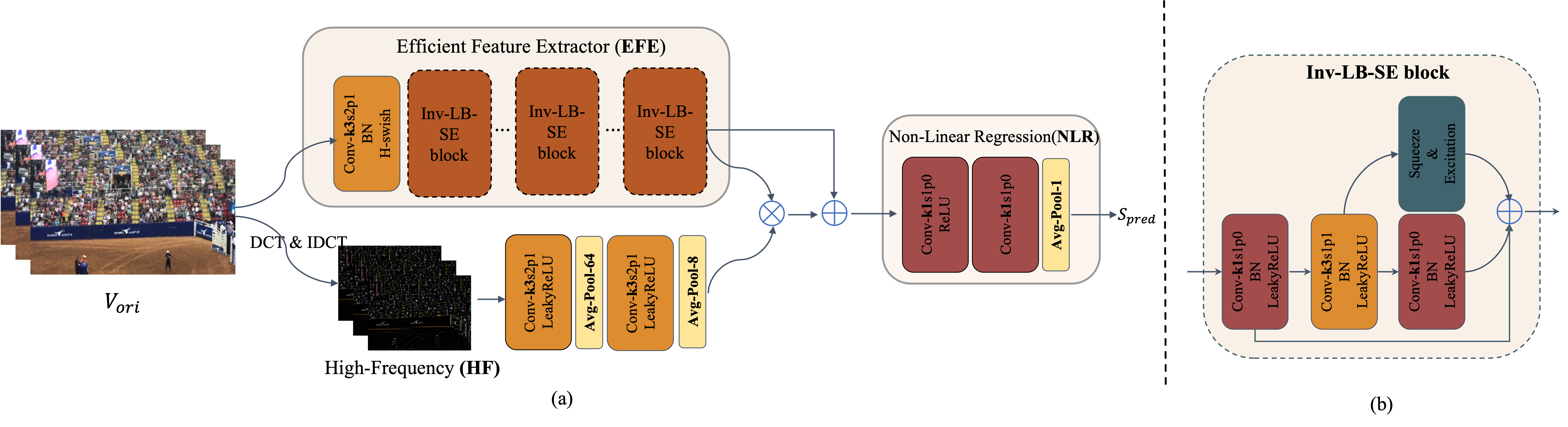}
\caption{(a) Overall framework for our proposed FreqSP.  (b) The detailed structure of the Inv-LB-SE block from ~\cite{howard2019searching}. Here we denote each convolutional layer with the corresponding kernel size (k), stride (s), and padding size (p).}
\label{fig:FreqSP}
\end{figure*}

To further illustrate the details, as shown in Fig.~\ref{fig:label}, we take a video randomly sampled from LSVQ~\cite{ying2021patch} as an example to clarify our labeling process. Given the uncompressed video, we first sharpen it at seven levels (0.0, 0.5, 1.0, 1.5, 2.0, 2.5, 3.0) using the built-in USM function of FFmpeg~\cite{jain1989fundamentals}, and then encode the sharpened videos using the HEVC/H.265 codec~\cite{sullivan2012overview} across five CRF values (21, 24, 27, 30, 33). We define the pre-sharpening encoding process as different encoders by sharpening levels and plot the RD curves for each encoder using the bitrate and VMAF. The overall seven curves are displayed in Fig.~\ref{fig:label} (a). Then we consider the encoder with sharpening level 0.0 as the anchor to calculate the BD-Rate of the other sharpening levels. 
As shown in Fig.~\ref{fig:label} (b), the BD-Rate of the encoder with sharpening level 0.5 is equal to the area of the shaded region, \ieno, $\text{BD-Rate}_{sharp0.5}=-0.3252$, which means that the encoder with sharpening level 0.5 needs $32.52\%$ fewer bits than the anchor (the encoder with sharpening level 0.0) to achieve comparable video quality. 
All BD-Rate values of encoders with different sharpening levels are shown in Fig.~\ref{fig:label} (c). The encoder with sharpening level 0.5 has the minimal BD-Rate value, so we label this example video as 0.5 in Fig.~\ref{fig:label}(d).

\subsection{Frequency-assisted sharpening level prediction}
\label{sec:pred}

Our Frequency-assisted Sharpening level Prediction model (FreqSP) is shown in Fig.~\ref{fig:FreqSP} (a). We indicate the convolutional layer by the corresponding kernel size (k), stride (s) and padding size (p). Taking the uncompressed raw video as input, we first learn intricate CNN features containing bitrate and quality information through the Efficient Feature Extractor (EFE). Meanwhile, we extract the High-Frequency components (HF) of the input video using the DCT $\&$ IDCT transform and feed them into two Conv-LeakyRelu-Pooling layers to get the filtered high-frequency features. Then, the filtered high-frequency information is fused with the output features of 15 Inv-LB-SE blocks using a residual connection and finally fed into a Non-Linear Regression head (NLR) to predict the final sharpening level.


\textbf{Efficient Feature Extractor (EFE)} Our FreqSP is only trained with uncompressed raw videos to predict sharpening levels with the optimal bitrate and quality tradeoff. To do this, the model must learn intricate features and compact representations of the original video, including those related to perceptual quality and video encoding. Inspired by the usage of Convolutional Neural Networks (CNN) in the field of video quality assessment~\cite{li2019quality} and video compression~\cite{li2021deep}, we employ convolutional blocks to learn hierarchical information connected to BD-Rate-related features. Instead of stacking deep and heavyweight convolutional blocks, we exploit computation-efficient CNN structures as our feature extraction branch, which is beneficial for practical deployment. To be specific, we utilize 15 inverted residual blocks with Linear-Bottleneck and Squeeze-and-Excitation attention (Inv-LB-SE blocks) from ~\cite{howard2019searching} to construct our Efficient Feature Extractor (EFE), where the first 3 Inv-LB-SE blocks have no SE shortcuts. The detailed Inv-LB-SE block is shown in Fig.~\ref{fig:FreqSP} (b), we replace ReLU with LeakyReLU to avoid the dying ReLU problem~\cite{lu2019dying}.

\textbf{High-Frequency (HF)}
On the other hand, the model is primarily predicting the sharpening level and the fundamental goal of the sharpening is to enhance the details and texture. After revisiting the sharpening algorithm Unsharp Masking (USM)~\cite{jain1989fundamentals} in FFmpeg, we attempt to utilize the high-frequency components extracted from the input raw video pertinent to the sharpening level to assist EFE for feature learning and result prediction. Specifically, USM sharpens image I by adding the detail layer $I_{D}$ to itself by a factor $\lambda$, 
\begin{equation}
\begin{aligned}
        &I_{usm} = I + \lambda I_{D},
        & I_{D} = I - I_{lp},
\end{aligned}
\end{equation}
where $I_{D}$ contains the high-frequency energy associated with fine details of $I$. And $I_{D}$ is generated by subtracting the original image $I$ to its low-pass filtered image $I_{lp}$. Sharpening, in other words, is the addition of non-low-frequency information to the original. As a result, we can acquire features that are closely related to sharpening by utilizing the high-frequency information in the original image. To extract high-frequency information, we first convert image $I^{C*H*W}$ from RGB to YCrCb and then adopt the discrete cosine transformation (DCT)~\cite{ahmed1974discrete} with block size $8\times8$ to get frequency maps $F^{64C*\frac{H}{8}*\frac{W}{8}}_{YCrCb}$, where each 64 frequency channels of Y channel, Cr channel and Cb channel is in order from low frequency to high frequency. After removing the first low-frequency channel in each of the 64 frequency channels and applying the inverse discrete cosine transform (IDCT), we get the corresponding high-frequency components as shown in Fig.~\ref{fig:FreqSP}.

\begin{table*}
\centering
\caption{Ablation studies on the prediction performance with different computation-efficient CNN structures of EFE on the LSVQ\cite{ying2021patch}, KoNViD-1k\cite{hosu2017konstanz} and LIVE-VQC\cite{sinno2018large} datasets.}
\label{table:cmp}
\begin{tabular}{|c|cc|cc|cc|}
\hline
& \multicolumn{2}{c|}{$\text{LSVQ}_{test}$ (2571)} & \multicolumn{2}{c|}{KoNViD-1k (1164)} & \multicolumn{2}{c|}{LIVE-VQC (585)} \\ \cline{2-7} 
\multirow{-2}{*}{Backbone} & PLCC ($\uparrow$)   & RMSE ($\downarrow$)  & PLCC ($\uparrow$)   & RMSE ($\downarrow$)  & PLCC ($\uparrow$)      & RMSE ($\downarrow$)     \\ \hline
Res-Layer (depth 18)                  & 0.7637            & 0.6597            & 0.8322                 & 0.5952                 & 0.7815                & 0.6413                \\
Res-Layer (depth 34)                  & 0.7691            & 0.6521            & 0.8358                 & 0.5887                 & 0.8084                & 0.6004                \\
Rep-Layer (depth 22)                  & 0.7454            & 0.6848            & 0.8166                 & 0.6223                 & 0.7812                & 0.6416                \\
Rep-Layer (depth 28)                  & 0.7450            & 0.6853            & 0.8363                 & 0.5879                 & 0.8023                & 0.6100                \\
\textbf{Inv-LB-SE  (depth 11) (Ours)} & 0.7583            & 0.6672            & 0.8447                 & 0.5726                 & 0.7967                & 0.6185                \\
\textbf{Inv-LB-SE (depth 15) (Ours)}  & \textbf{0.7716 }           & \textbf{0.6486}            & \textbf{0.8601}        & \textbf{0.5434}        & \textbf{0.8116}       & \textbf{0.5955}     \\
\hline
\end{tabular}
\end{table*}

\begin{table*}
\centering
\caption{Speed test of our proposed EFE with different computation-efficient CNN structures on GPU (A100-SXM-80GB) and CPU (Intel-Xeon-Platinum-8336C-CPU). The results of time usage are average of 20 runs after warming up the hardware with a single thread.}
\label{table:speed}
\begin{tabular}{|c|c|c|c|cc|}
\hline
\multirow{2}{*}{Backbone} & \multirow{2}{*}{Params/M} & \multirow{2}{*}{Memory/M} & \multirow{2}{*}{FLOPs/G} & \multicolumn{2}{c|}{Runtime/ms} \\ \cline{5-6} &     &               &    & $cpu_{t1}$           & $gpu_{t1}$       \\ \hline
Res-Layer (depth 18)         & 12.89                                         & 1198.45                                       & 118.7                                        & 3810.7104          & \textbf{8.5665} \\
Res-Layer (depth 34)         & 23                                            & 1277.39                                       & 196.2                                        & 5579.6474          & 11.5169         \\
Rep-Layer (depth 22)         & 12.1                                          & 1199.28                                       & 164.4                                        & 4523.5404          & 13.9992         \\
Rep-Layer (depth 28)         & 18.81                                         & 1253.24                                       & 243.8                                        & 7046.9689          & 16.3129         \\
\textbf{Inv-LB-SE  (depth 11) (Ours)} & \textbf{2.85}                                 & \textbf{1135.19}                              & \textbf{49.9}                                & \textbf{2146.5186} & 10.422          \\
\textbf{Inv-LB-SE (depth 15) (Ours) } & 6.18                                          & 1170.13                                       & 86                                           & 3604.8783          & 11.8494        \\
\hline
\end{tabular}
\end{table*}

\textbf{Non-Linear Regression (NLR)} 
We exploit non-linear $1\times1$ convolutional layer~\cite{szegedy2015going} with forms of $k1s1p0$ to replace commonly used linear fully-connected layer (FC) in video recognition~\cite{Wang_2018_CVPR,Chen_2018_ECCV} to achieve dimension reduction by decreasing the number of feature maps whilst retaining their salient features. To avoid losing the sensitivity to the diverse information of BD-Rate-related features fused with HF, we average them only after the dimension reduction and put Average Pooling (Avg-Pool) as the last regression layer to output the final predicted sharpening level.

\subsection{Objective Functions}
\label{sec:loss}

We define the overall loss function as the weighted sum of
monotonicity loss $\mathcal{L}_{\text{mono}}$ and L1 loss $\mathcal{L}_{\text{L1}}$ as follows:
\begin{equation}
    \begin{aligned}
        & \mathcal{L}_{\text{mono}} = \sum_{i,j}max((S_{pred}^{i}, S_{pred}^{j})sgn(S_{gt}^{j}-S_{gt}^{i}), 0),\\
        & \mathcal{L}_{\text{overall}} = \mathcal{L}_{\text{L1}}(S_{pred}, S_{gt}) + \lambda \mathcal{L}_{\text{mono}} ,
    \end{aligned}
\end{equation}
where $sgn(\cdot)$ denotes the sign function. $S_{pred}$ and $S_{gt}$ refer to predicted results and ground truth respectively.

\section{Experiments}
\label{sec:exps}
In this section, we first discuss the datasets utilized for training and testing. We then provide an in-depth overview of our experimental setup, outlining the hyper-parameters and metrics employed. To the best of our knowledge, there is no related work concerning our proposed problem. Hence, we conduct several ablation studies to compare and analyze the performances of our approach.


\subsection{Dataset}
We choose 12854 videos from large-scale LSVQ\cite{ying2021patch} for training and testing by the ratio 8:2. To further demonstrate the generalization ability of our proposed model, we directly perform cross-dataset evaluations on two widely-recognized in-the-wild natural video benchmark datasets, KoNViD-1k~\cite{hosu2017konstanz} with 1164 videos and LIVE-VQC~\cite{sinno2018large} with 585 videos, respectively.

We extract 32 frames from each video during the training and testing phases and crop them for data augmentation. Instead of utilizing typical random cropping or center cropping, we divide each frame into several $16\times16$ patches and cut each patch into multiple $16\times16$ blocks, training with random position and testing with the fixed top left corner. We then re-stitch these blocks into a $256\times256$ size image according to the original position of the patches, where the re-stitched image is used as the input of our model.

\begin{table*}
\centering
\caption{Ablation studies on Efficient Feature Extractor (EFE), Non-Linear Regression (NLR) and High Frequency (HF).}
\label{table:cmp2}
\begin{tabular}{|c|cc|cc|cc|}
\hline
& \multicolumn{2}{c|}{$\text{LSVQ}_{test}$ (2571)} & \multicolumn{2}{c|}{KoNViD-1k (1164)} & \multicolumn{2}{c|}{LIVE-VQC (585)} \\ \cline{2-7} 
\multirow{-2}{*}{Inv-LB-SE (depth 15)} & PLCC ($\uparrow$)   & RMSE ($\downarrow$)  & PLCC ($\uparrow$)   & RMSE ($\downarrow$)  & PLCC ($\uparrow$)      & RMSE ($\downarrow$)     \\ \hline
NLR + w/o HF                   & 0.7680    & 0.6537   & 0.8400     & 0.5812      & 0.7878      & 0.6320         \\
LR + HF                        & 0.7642     & 0.6590    & 0.8403     & 0.5807     & 0.8114     & 0.5958                \\
NLR + w/o EFE                  & 0.5924    & 0.8664      & 0.7043    & 0.7902   & 0.6611      & 0.7986                \\
LR + w/o EFE                   & 0.5683    & 0.8917    & 0.6151     & 0.9015    & 0.5275      & 0.9430                \\
LR + w/o HF                    & 0.7636    & 0.6598     & 0.8295    & 0.6000     & 0.7941     & 0.6224                \\
\textbf{NLR + HF (ours)}      & \textbf{0.7716}    & \textbf{0.6486}     & \textbf{0.8601}    & \textbf{0.5434} & \textbf{0.8116}      & \textbf{0.5955}       \\    \hline
\end{tabular}
\end{table*}

\begin{figure*}[!t]
\centering
\includegraphics[width=\textwidth]{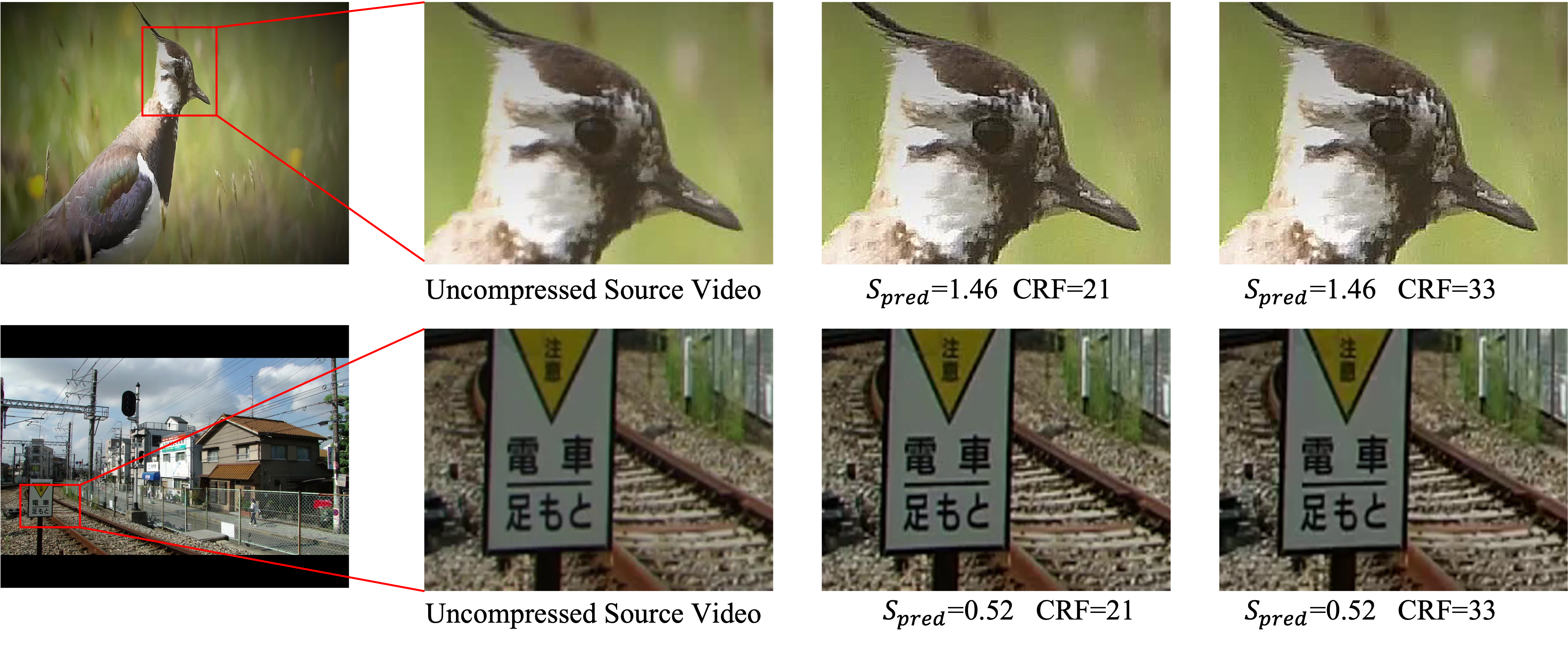}
\caption{Visualization of sharpened video frames with the predicted sharpening level from our model.}
\label{fig:vis}
\end{figure*}

\subsection{Implementation Details}
We train our model using the AdamW optimizer~\cite{DBLP:conf/iclr/LoshchilovH19} with an initial learning rate $0.001$ and weight decay $0.05$. We load the weights of each Inv-LB-SE block of EFE from the matching layers of Mobilenetv3 trained on ImageNet dataset~\cite{russakovsky2015imagenet} as our initial training states. Generally, the weight of monotonicity loss $\mathcal{L}_{\text{mono}}$ is set to $\lambda=0.3$. We set the batch size to 16. We use PLCC (Pearson linear correlation coefficient) and RMSE as metrics. Our model is implemented based on the PyTorch framework with a single NVIDIA A100-SXM-80GB GPU. 

\subsection{Experimental Results}
Since there are no corresponding baselines, we conduct ablation studies on the prediction performance and testing speed with various types of computation-efficient CNN structures of EFE. Moreover, we investigate the different properties of our model to illustrate the role of each designed component.

\begin{figure*}[!t]
\centering
\includegraphics[width=\textwidth]{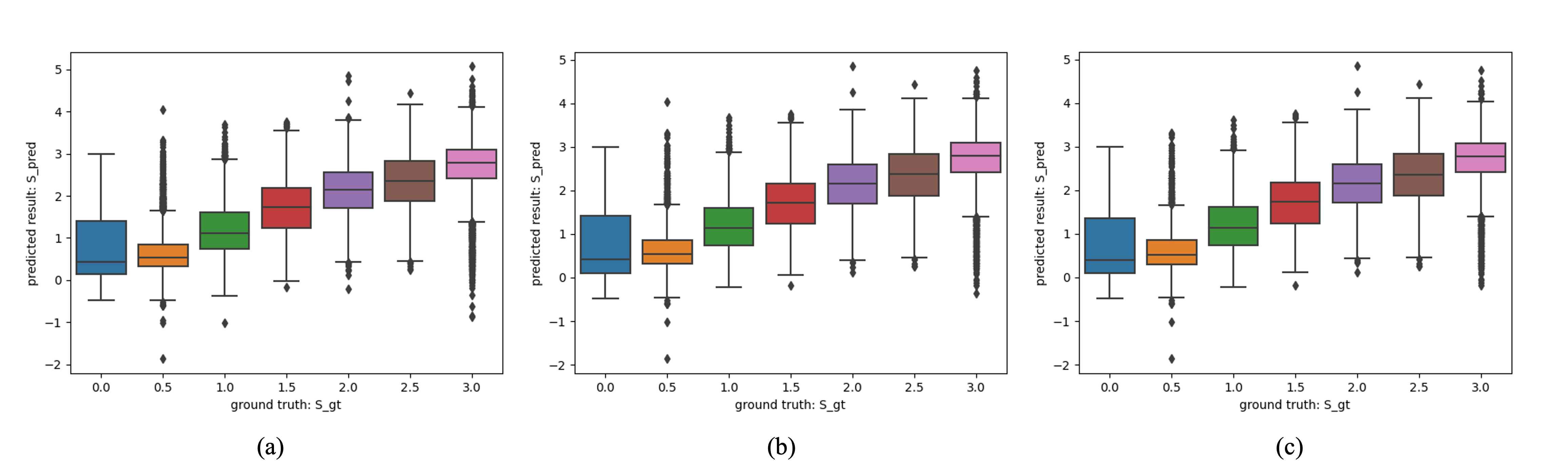}
\caption{Relationship between pseudo-label $S_{gt}$ and our predicted results $S_{pred}$ of Inv-LB-SE (depth 15) on testing datasets (a) $\text{LSVQ}_{test}$, (b) KoNViD-1k, (c) LIVE-VQC, respectively. }
\label{fig:pred}
\end{figure*}

To validate the prediction performance, we replace the convolutional layers in EFE with various computationally efficient CNN structures on three datasets LSVQ\cite{ying2021patch}, KoNViD-1k\cite{hosu2017konstanz} and LIVE-VQC\cite{sinno2018large}. In detail, we test the Inv-LB-SE block from MobilenetV3~\cite{howard2019searching} with depths 11 and 15, residual layer (Res-Layer) from ResNet~\cite{he2016deep} with depths 18 and 34 and re-parameterized VGG layer (Rep-Layer) from RepVGG~\cite{ding2021repvgg} with depths 22 and 28. 
As shown in Table.~\ref{table:cmp}, for each efficient CNN layer, as the number of network layers rises (\ieno, the deeper), the better the prediction performance. 
Generally, our EFE with 15 Inv-LB-SE layers achieves the best PLCC and RMSE scores on LSVQ, KoNViD-1k and LIVE-VQC. 

As shown in Fig.~\ref{fig:vis}, we visualize the pre-sharpening compressed frames using our predicted sharpening level $S_{pred}$ across CRF 21, 33. All results are sharper and more aesthetically pleasing than the source frames.
To intuitively illustrate the prediction accuracy of our model, as shown in Fig.~\ref{fig:pred}, we visualize the correlation of ground truth $S_{gt}$ and predicted results $S_{pred}$ of Inv-LB-SE (depth 15) on testing datasets (a) $\text{LSVQ}_{test}$, (b) KoNViD-1k and (c) LIVE-VQC. 

To evaluate the performance of our model in real industrial deployment, we measure the inference speed with various computationally efficient CNN structures of EFE on GPU (A100-SXM-80GB) and CPU (Intel-Xeon-Platinum-8336C-CPU), including the parameters (Params), Memory, FLOPs and the actual running time on CPU ($cpu_{t1}$) and GPU ($gpu_{t1}$). The results of time usage are average of 20 runs after warming up the hardware with a single thread ($t1$). As shown in Table.~\ref{table:speed}, our EFE with 11 Inv-LB-SE blocks has the fewest parameters, lowest memory usage and the fastest CPU inference speed. Our EFE with 18 Res-Layer has the fastest GPU inference speed.

Moreover, we also conduct ablation studies on each designed component. As shown in Table.~\ref{table:cmp2}, our proposed EFE contributes to the performance by up to almost $0.2$ RMSE decreases on three datasets. The proposed HF module could bring notable RMSE improvements on $\text{LSVQ}_{test} (-0.780\%)$ , KoNViD-1k ($-6.504\%$) and LIVE-VQC ($-5.775\%$). When we replace our NLR module with a general linear regression (LR) to implement prediction, we see that our NLR module outperforms LR with non-negligible improvements on three datasets.

\section{Conclusion}

In this paper, we present a Frequency-assisted Sharpening level Prediction model (FreqSP) that utilizes uncompressed source videos to predict the optimal sharpening level considering the bitrate and quality tradeoff. We first pseudo-label each training video with the sharpening level deriving from its BD-Rate characteristics as ground truth. Then we propose FreqSP
by designing the EFE module to learn intricate features and extracting high-frequency features to assist the sharpening level prediction. We also propose non-linear regression to retain the most important features and estimate final prediction results. Extensive experimental results have shown the effectiveness of our method.

\bibliographystyle{IEEEbib}
\bibliography{main}

\begin{thebibliography}{10}

\bibitem{ying2021patch}
Zhenqiang Ying, Maniratnam Mandal, Deepti Ghadiyaram, and Alan Bovik,
\newblock ``Patch-vq:'patching up'the video quality problem,''
\newblock in {\em Proceedings of the IEEE/CVF Conference on Computer Vision and Pattern Recognition}, 2021, pp. 14019--14029.

\bibitem{jain1989fundamentals}
Anil~K Jain,
\newblock {\em Fundamentals of digital image processing},
\newblock Prentice-Hall, Inc., 1989.

\bibitem{sullivan2012overview}
Gary~J Sullivan, Jens-Rainer Ohm, Woo-Jin Han, and Thomas Wiegand,
\newblock ``Overview of the high efficiency video coding (hevc) standard,''
\newblock {\em IEEE Transactions on circuits and systems for video technology}, vol. 22, no. 12, pp. 1649--1668, 2012.

\bibitem{bjontegaard2001calculation}
Gisle Bjontegaard,
\newblock ``Calculation of average psnr differences between rd-curves,''
\newblock {\em VCEG-M33}, 2001.

\bibitem{howard2019searching}
Andrew Howard, Mark Sandler, Grace Chu, Liang-Chieh Chen, Bo~Chen, Mingxing Tan, Weijun Wang, Yukun Zhu, Ruoming Pang, Vijay Vasudevan, et~al.,
\newblock ``Searching for mobilenetv3,''
\newblock in {\em Proceedings of the IEEE/CVF international conference on computer vision}, 2019, pp. 1314--1324.

\bibitem{li2019quality}
Dingquan Li, Tingting Jiang, and Ming Jiang,
\newblock ``Quality assessment of in-the-wild videos,''
\newblock in {\em Proceedings of the 27th ACM International Conference on Multimedia}, 2019, pp. 2351--2359.

\bibitem{li2021deep}
Jiahao Li, Bin Li, and Yan Lu,
\newblock ``Deep contextual video compression,''
\newblock {\em Advances in Neural Information Processing Systems}, vol. 34, pp. 18114--18125, 2021.

\bibitem{lu2019dying}
Lu~Lu, Yeonjong Shin, Yanhui Su, and George~Em Karniadakis,
\newblock ``Dying relu and initialization: Theory and numerical examples,'' 2019.

\bibitem{ahmed1974discrete}
Nasir Ahmed, T\_ Natarajan, and Kamisetty~R Rao,
\newblock ``Discrete cosine transform,''
\newblock {\em IEEE transactions on Computers}, vol. 100, no. 1, pp. 90--93, 1974.

\bibitem{hosu2017konstanz}
Vlad Hosu, Franz Hahn, Mohsen Jenadeleh, Hanhe Lin, Hui Men, Tam{\'a}s Szir{\'a}nyi, Shujun Li, and Dietmar Saupe,
\newblock ``The konstanz natural video database (konvid-1k),''
\newblock in {\em 2017 Ninth international conference on quality of multimedia experience (QoMEX)}. IEEE, 2017, pp. 1--6.

\bibitem{sinno2018large}
Zeina Sinno and Alan~Conrad Bovik,
\newblock ``Large-scale study of perceptual video quality,''
\newblock {\em IEEE Transactions on Image Processing}, vol. 28, no. 2, pp. 612--627, 2018.

\bibitem{szegedy2015going}
Christian Szegedy, Wei Liu, Yangqing Jia, Pierre Sermanet, Scott Reed, Dragomir Anguelov, Dumitru Erhan, Vincent Vanhoucke, and Andrew Rabinovich,
\newblock ``Going deeper with convolutions,''
\newblock in {\em Proceedings of the IEEE conference on computer vision and pattern recognition}, 2015, pp. 1--9.

\bibitem{Wang_2018_CVPR}
Limin Wang, Wei Li, Wen Li, and Luc Van~Gool,
\newblock ``Appearance-and-relation networks for video classification,''
\newblock in {\em Proceedings of the IEEE Conference on Computer Vision and Pattern Recognition (CVPR)}, June 2018.

\bibitem{Chen_2018_ECCV}
Yunpeng Chen, Yannis Kalantidis, Jianshu Li, Shuicheng Yan, and Jiashi Feng,
\newblock ``Multi-fiber networks for video recognition,''
\newblock in {\em Proceedings of the European Conference on Computer Vision (ECCV)}, September 2018.

\bibitem{DBLP:conf/iclr/LoshchilovH19}
Ilya Loshchilov and Frank Hutter,
\newblock ``Decoupled weight decay regularization,''
\newblock in {\em International Conference on Learning Representations {ICLR}}, 2019.

\bibitem{russakovsky2015imagenet}
Olga Russakovsky, Jia Deng, Hao Su, Jonathan Krause, Sanjeev Satheesh, Sean Ma, Zhiheng Huang, Andrej Karpathy, Aditya Khosla, Michael Bernstein, et~al.,
\newblock ``Imagenet large scale visual recognition challenge,''
\newblock {\em International journal of computer vision}, vol. 115, no. 3, pp. 211--252, 2015.

\bibitem{he2016deep}
Kaiming He, Xiangyu Zhang, Shaoqing Ren, and Jian Sun,
\newblock ``Deep residual learning for image recognition,''
\newblock in {\em Proceedings of the IEEE conference on computer vision and pattern recognition}, 2016, pp. 770--778.

\bibitem{ding2021repvgg}
Xiaohan Ding, Xiangyu Zhang, Ningning Ma, Jungong Han, Guiguang Ding, and Jian Sun,
\newblock ``Repvgg: Making vgg-style convnets great again,''
\newblock in {\em Proceedings of the IEEE/CVF Conference on Computer Vision and Pattern Recognition}, 2021, pp. 13733--13742.

\end{thebibliography}

\end{document}